\documentclass[12pt]{article}
\def\lae{\;^{<}_{\sim} \;} \def\gae{\; ^{>}_{\sim} \;}

\title{\textbf{Large non-Gaussianity generated at the end of Extended D-term Hybrid Inflation}}
{\author{\\[1cm]
{\sc \large Chia-Min Lin$^\ast$}\\[1cm]
{\sl\small Department of Physics, National Tsing Hua University, Hsinchu, Taiwan 300 }\\[1cm]
}}

\usepackage[margin=2cm]{geometry}
\usepackage{graphicx,color}
\usepackage{subfigure}
\begin{document}
\maketitle
\begin{abstract}
In this paper, we show that if we extend D-term hybrid inflation by adding a light scalar field which couples to a waterfall field, large non-Gaussianity can be generated at the end of inflation. Contribution of cosmic strings generated after D-term inflation to the Cosmic Microwave Background (CMB) angular power spectrum can be as low as $10\%$ and this also evade the spectral index problem.
\end{abstract}
\footnoterule{\small $^\ast$cmlin@phys.nthu.edu.tw}
\section{Introduction}
Inflation is interesting because by assuming an accelerating phase of the early universe, we can solve many problems for the hot big bang model. One of an interesting model of inflation is D-term hybrid inflation \cite{Binetruy:1996xj, Halyo:1996pp, Riotto:1997wy, Lyth:1997pf}. D-term inflation not only is a supersymmetric inflation model but can also be derived from string theory known as D3/D7 brane inflation \cite{Dasgupta:2002ew, Koyama:2003yc}. However, for conventional D-term hybrid inflation without tuning the coupling constants, there are cosmic string problem and spectral index problem. The first problem is due to the fact that if the curvature perturbation is generated by the inflaton field, then it will fix the scale of inflation to be too high so that the energy density of cosmic strings generated after inflation is larger than allowed from current observation. The second problem is conventional D-term inflation predicts the spectral index $n_s \gae 0.98$ while latest WMAP result suggest $n_s \simeq 0.96$ \cite{Komatsu:2008hk}. However, if we allow about $10\%$ contribution to the CMB angular power spectrum from cosmic strings, the favored spectral index becomes $n_s \simeq 1$ \cite{Bevis:2007gh, Battye:2006pk}. Many papers are dealing with the cosmic string problem and/or the spectral index problem, see for example \cite{Endo:2003fr, Seto:2005qg, Urrestilla:2004eh, Rocher:2004my, Lin:2006xta, Lin:2007va, Lin:2008ys}.

However, if large non-Gaussianity in the CMB is detected in the future, then D-term inflation has to be abandoned or to be extended in such a way that non-Gaussianity can be generated. As an example, see \cite{Bernardeau:2007xi}. In \cite{Lin:2009yn}, it is shown that by introducing a right-handed sneutrino curvaton, large non-Gaussianity can be generated and also the problems of cosmic strings and spectral index can be evaded. In this paper, we propose another method to achieve this goal. We consider the case that the curvature perturbation is generated at the end of inflation by the method introduced in \cite{Lyth:2005qk}.

This paper is organized as follows. In section \ref{2}, we present the basic equations we are going to use. In section \ref{33}, we review D-term inflation. In section \ref{4}, we apply the method of generating curvature perturbation at the end of inflation to D-term inflation. In section \ref{3}, we calculate the amount of non-Gaussianity which can be generated in this framework. In section \ref{6}, we discuss the effect of supergravity and show that it is negligible. Section \ref{c} is our conclusion.
\section{Generating the curvature perturbation at the end of inflation}
\label{2}

In this section, we briefly review the method introduced in \cite{Lyth:2005qk} in order to present the formulas we are going to apply in the next section.

From $\delta N$ formalism \cite{Lyth:2005fi, Sasaki:1995aw}, we know that if a field $\phi$ is relevant for generating the curvature perturbation $\zeta$, then
\begin{equation}
\zeta=\delta N=N_\phi \delta \phi+\frac{1}{2}N_{\phi\phi}(\delta \phi)^2,
\end{equation}
where $N_\phi \equiv \partial N/\partial \phi$ and $N_{\phi\phi} \equiv \partial^2 N/\partial \phi^2$. For single field slow-roll inflation with inflaton $s$ and its scalar potential $V$, the number of e-folds $N$ is related to $s$ via \cite{Lyth:1998xn}
\begin{equation}
dN=\frac{1}{M_P^2}\frac{V}{V^\prime}ds=\frac{1}{\sqrt{2\epsilon}M_P}ds,
\label{eq2}
\end{equation}
where $V^\prime \equiv dV/ds$ and $\epsilon \equiv (M_P^2/2)(V^\prime/V)^2$ is a slow-roll parameter. Another slow-roll parameter is $\eta \equiv M_P^2 V^{\prime\prime}/V$. During slow-roll inflation both $\epsilon$ and $\eta$ are much smaller than 1. Inflation ends if the absolute value of either one of the slow-roll parameters becomes 1. If the curvature perturbation is generated during inflation, we have
\begin{equation}
N_s=\frac{1}{\sqrt{2}\epsilon M_P}
\end{equation}
and the spectrum $P_\zeta\sim \zeta^2$ is
\begin{equation}
P_\zeta \simeq N_s^2 (\delta s)^2 =\frac{H^2}{8\pi^2\epsilon M_P^2}.
\label{eq4}
\end{equation}
This is a standard result of single field slow-roll inflation where we have used the fact that $\delta s \sim H/2\pi$. From observation of Cosmic Microwave Background (CMB), we know that $P_\zeta^{1/2} \simeq 5 \times 10^{-5}$ \cite{Komatsu:2008hk}. We will refer to this as CMB normalization.

The idea of generating the curvature perturbation at the end of inflation is that at the end of inflation the inflaton field value $s_e$\footnote{We use a subscript e to denote the value at the end of inflation (N=0) and a subscript N to denote the value at number of e-folds N throughout this paper.} depends on another light scalar field $\phi$ so that the quantum fluctuation of $\phi$ during inflation will result in a fluctuation of $s_e$ thus a curvature perturbation $\zeta=\delta N$. From Eq. (\ref{eq2}), near the end of inflation
\begin{equation}
dN=\frac{1}{\sqrt{2\epsilon_e}M_P}ds_e=\frac{1}{\sqrt{2\epsilon_e}M_P}s_e^\prime d\phi,
\end{equation}
where $s_e^\prime \equiv ds_e/d\phi$. Therefore we get
\begin{equation}
N_\phi=\frac{1}{\sqrt{2\epsilon_e}M_P}s_e^\prime \;\;\; \mbox{and} \;\;\; N_{\phi\phi}=\frac{1}{\sqrt{2\epsilon_e}M_P}s_e^{\prime\prime}
\end{equation}
thus the spectrum generated at the end of inflation is
\begin{equation}
P_\zeta \simeq N^2_\phi (\delta \phi)^2=\frac{H^2}{8\pi^2 \epsilon_e M_P^2}(s^\prime_e)^2
\label{eq7}
\end{equation}
where we have used the fact that $\delta \phi \sim H/2\pi$. The amount of non-Gaussianity is characterized by the non-linear parameter $f_{NL}$ which is given by
\begin{equation}
f_{NL}=-\frac{5}{6}\frac{N_{\phi\phi}}{N_\phi^2}=-\frac{5}{6}\frac{\sqrt{2\epsilon_e}s_e^{\prime\prime}}{(s_e^\prime)^2}.
\label{eq8}
\end{equation}

In this paper, we assume that the curvature perturbation is dominated by those generated at the end of inflation. Comparing Eq. (\ref{eq4}) and (\ref{eq7}), this implies
\begin{equation}
(s_e^\prime)^2 \gg \frac{\epsilon_e}{\epsilon}.
\label{eq9}
\end{equation}
\section{D-term Inflation}
\label{33}
The superpotential of D-term hybrid inflation is given by
\begin{equation}
W=\lambda S \Phi_+ \Phi_-,
\end{equation}
where $S$ is the inflaton superfield, $\lambda$ is the superpotential coupling and $\Phi_\pm$ are chiral superfields charged under the $U(1)_{FI}$ gauge symmetry responsible for the Fayet-Iliopoulos term. The corresponding scalar potential is
\begin{equation}
V(S, \Phi_+, \Phi_-)=\lambda^2\left[|S|^2(|\Phi_+|^2+|\Phi_-|^2)+|\Phi_+|^2|\Phi_-|^2\right]+\frac{g^2}{2}\left(|\Phi_+|^2-|\Phi_-|^2+\xi\right)^2,
\label{eq11}
\end{equation}
where $\xi$ is the Fayet-Iliopoulos term and $g$ is the $U(1)_{FI}$
gauge coupling. A very small $g$ (far smaller than order $O(1)$) is
regarded as unnatural, because we do not know of any small (for
example $g<10^{-3}$) gauge couplings in particle physics. In this paper, we choose $g=0.1$. The true
vacuum of the potential is given by $|S|=|\Phi_+|=0$ and
$|\Phi_-|=\sqrt{\xi}$.  When $|S| > |S_c|= g \xi^{1/2}/\lambda$,
there is a local minimum occurred at $|\Phi_+|=|\Phi_-|=0$.
Therefore,  at tree level the potential is just a constant $V=g^2
\xi^2/2$. The 1-loop corrections to $V$ can be calculated using the
Coleman-Weinberg formula \cite{Coleman:1973jx}
\begin{equation}
\Delta V=\frac{1}{64\pi^2}\sum_i(-1)^F m^4_i\ln\frac{m^2_i}{\Lambda^2},
\end{equation}
with $m_i$ being the mass of a given particle, where the sum goes
over all particles with $F=0$ for bosons and $F=1$ for fermions and
$\Lambda$ is a renormalization scale. Thus the 1-loop potential is
given by (setting $s=\sqrt{2}\mbox{Re}(S)$)
\begin{equation}
V(s)=V_0+\frac{g^4 \xi^2}{16 \pi^2}\ln \left(\frac{s^2}{2\Lambda^2}\right),
\end{equation}
where $V_0=g^2\xi^2/2$. In this paper, we set $\xi^{1/2}=1.67 \times 10^{-3}M_P$ which corresponds to about $10\%$ contribution from cosmic strings generated after D-term inflation to CMB.

Using $s_c=\sqrt{2}g\xi^{1/2}/\lambda$, by integrate Eq. (\ref{eq2}), we can obtain $s^2_N=g^2 N M_P^2/2\pi^2+s^2_c$. Thus the slow-roll parameter $\epsilon$ at the end of inflation and during inflation are given by
\begin{equation}
\epsilon_e =\frac{g^2 \lambda^2}{64 \pi^4 \xi} \;\;\;\; \mbox{and} \;\;\;\; \epsilon_N=\frac{g^4}{32 \pi^4 \left(\frac{2 g^2 \xi}{\lambda^2}+\frac{g^2 N M_P^2}{2 \pi^2}\right)}
\label{epsilone}
\end{equation}
respectively.

We require that inflation ends via waterfall field before the fail of slow-roll condition $|\eta|=1$, this implies.
\begin{equation}
\frac{\sqrt{2}g \xi^{1/2}}{\lambda} > \frac{g}{2 \pi}.
\end{equation}
Hence $\lambda< 1.48 \times 10^{-2}$.
We also require that the fields in our model are less than Planck mass, including $s_c$, i.e.
\begin{equation}
\frac{\sqrt{2} g \xi^{1/2}}{\lambda}<M_P.
\end{equation}
Hence $\lambda > 2.36 \times 10^{-4}$. Therefore
\begin{equation}
2.36\times 10^{-4}<\lambda<1.48\times 10^{-2}.
\label{eq17}
\end{equation}
\section{Generating curvature perturbation at the end of extended D-term inflation}
\label{4}
In this paper, we extend D-term inflation by adding a light scalar field $\Phi$ coupled to a waterfall field $\Phi_-$ via $x|\Phi|^2|\Phi_-|^2$.
Instead of Eq. (\ref{eq11}), we now have
\begin{equation}
V \supset \lambda^2 \left(|S|^2+\frac{x}{\lambda^2}|\Phi|^2-|S_c|^2\right)|\Phi_-|^2
\end{equation}
for the mass term of $\Phi_-$, by setting $\phi=\sqrt{2}\mbox{Re}(\Phi)$, we obtain
\begin{equation}
s_e^2=s_c^2-\frac{x}{\lambda^2}\phi^2
\label{eq19}
\end{equation}
hence
\begin{equation}
s_e^\prime=-\frac{x\phi}{\lambda^2 s_e}.
\end{equation}
Here we require that $s_e^2>0$. This puts an upper bound for $\phi$, namely, $\phi^2<2g^2\xi/x$ which implies $\phi < 7.47 \times 10^{-4}M_P$ in our setup. A lower bound of $\phi$ can be obtained by requiring its classical value to be larger than its quantum fluctuation, that is, $\phi>H/2\pi=1.81 \times 10^{-8}M_P$ in our setup. Therefore we have
\begin{equation}
1.81 \times 10^{-8}M_P< \phi < 7.47 \times 10^{-4}M_P
\label{constraint2}
\end{equation}

From Eq. (\ref{eq7}) and imposing CMB normalization, we obtain
\begin{equation}
P_\zeta=\frac{2 \xi^2 \pi^2 x^2 \phi^2}{3 g^2 \lambda^4}=2.5 \times 10^{-9},
\label{cmb}
\end{equation}
where we have assumed that $s_e \simeq s_c$ which can be justified by putting our values into Eq. (\ref{eq19}). For $x=0.1$ and $\phi=10^{-4}M_P$ we obtain $\lambda=3.78 \times 10^{-3}$. Note that this value satisfies the constraint Eq. (\ref{eq17}).

The spectral index is \cite{Lyth:2005qk}
\begin{equation}
n_s=1+2\eta_{\phi\phi}-2\epsilon_N,
\end{equation}
where $\eta_{\phi\phi} \equiv M_P^2(\partial^2 V/\partial
\phi^2)/V$. Because by assumption $\phi$ is a light field and
$\epsilon_N$ is found to be very small, we have $n_s \simeq 1$. This
is a favored result because as mentioned in the last section we have
about $10\%$ cosmic strings contribution to the CMB. We also have to check that whether Eq. (\ref{eq9}) is satisfied. Using the above results, we can see the l.h.s. of Eq. (\ref{eq9}) is $128$ and the r.h.s. is $8.78$ by using $N=60$. Therefore our
assumption is justified.

\section{Non-Gaussianity}
\label{3}

Currently the upper bound of $f_{NL}$ \cite{Yadav:2007yy, Komatsu:2008hk, Curto:2009pv} is roughly
\begin{equation}
f_{NL} \lae 100.
\end{equation}
In the near future, the Planck satellite\cite{:2006uk} will reduce the bound to $f_{NL} \lae 10$ if non-Gaussianity is not detected. Therefore, we investigate the possibility of producing $10<f_{NL}<100$ in this work. We refer this range as large non-Gaussianity.

By using Eq. (\ref{eq8}) and (\ref{cmb}), we can obtain
\begin{equation}
f_{NL}=\frac{4.3\times 10^{-4}}{\lambda^2}.
\end{equation}
For $f_{NL}=10$, we have $\lambda=6.56 \times 10^{-3}$ and $\phi=3.01 \times 10^{-4}$. For $f_{NL}=100$, we have $\lambda=2.07 \times 10^{-3}$ and $\phi=3.00 \times 10^{-5}$. Note that constaints from both Eq. (\ref{eq17}) and Eq. (\ref{constraint2}) are satisfied.

\section{Effects of Supergravity}
\label{6}
If we consider a light scalar field $\Phi$ with mass $m_\Phi$ couple to the waterfall field $\Phi_-$ through the K\"ahler potential of this form
\begin{equation}
K=\frac{c}{M_P^2}|\Phi_-|^2 |\Phi|^2,
\end{equation}
where $c$ is of $O(1)$. The appearance of this term cannot be prevented by any symmetry. From the F-term of supergravity, there is an additional contribution
to the scalar potential
\begin{equation}
\delta V=\frac{(1-c)m^2_{\Phi}}{M_P^2}|\Phi|^2|\Phi_-|^2.
\label{eq27}
\end{equation}
Hence we have a form similar to what we have in section \ref{4} with $x=(1-c)m^2_{\phi}/M_P^2$. A similar mechanism was used in \cite{Lin:2006xta, Lin:2007va}, but the purpose for those papers was to convert D-term inflation into hilltop inflation \cite{Kohri:2007gq, Boubekeur:2005zm}.

The Hubble parameter during D-term inflation is $H \lae 10^{-7}M_P^2$. By the assumption that $\phi$ is some light field, which means $m_{\phi} \ll H$. We can see from Eq. (\ref{eq27}) that $x$ is extremely small ($<O(10^{-14})$) in this case. Therefore we can see that supergravity effect in our model is negligible. Actually if supergravity effect is significant, then it should be taken seriously for many models based on a similar setting.

\section{Conclusions}
\label{c}
In this paper, we have shown that if we extend conventional D-term hybrid inflation by adding a light field which is coupled to a waterfall field. Then it is possible to reduce the scale of inflation thus evade too much cosmic strings produced after inflation, because the mechanism of producing curvature perturbation is now different. We have shown that large non-Gaussianity can be generated in this setup. We also allow about $10\%$ cosmic strings produced after inflation and take the advantage that $n_s=1$ is the favored value of the spectral index. It is also possible that cosmic strings can be detected in the future therefore out scenario can also be tested in this way.

\section*{Acknowledgement}
This work was supported in part by the
NSC under grant No. NSC 96-2628-M-007-002-MY3, by the NCTS, and by the
Boost Program of NTHU.

\end{document}